\documentclass[aps,prl,twocolumn,superscriptaddress]{revtex4-2} 
\usepackage[english]{babel}
\usepackage{amsmath, bm, amssymb}
\usepackage{mathtools}
\usepackage{siunitx}
\usepackage{environ}
\usepackage{mathrsfs}
\usepackage{braket}
\usepackage[]{hyperref}
\DeclareMathAlphabet{\mathscrlower}{OT1}{pzc}{m}{it} 
\usepackage{prettyref}
\usepackage{graphicx}
\usepackage{booktabs}
\usepackage{array}
\usepackage{multirow}
\usepackage{dcolumn}
\usepackage{threeparttable}
\usepackage{threeparttablex}
\usepackage{placeins}
\usepackage[stable]{footmisc}
\usepackage[version=4]{mhchem}
\usepackage{xfrac}
\usepackage{color}
\usepackage{xcolor}
\usepackage[normalem]{ulem}

\DeclareSIUnit\bohr{\text {\ensuremath {a}}_{0}}

\newcommand{\pauli}{\boldsymbol{\sigma}}

\newcommand{\diraccontra}[1]{\boldsymbol{\gamma}^{#1}}

\newcommand{\diraca}{\vec{\boldsymbol{\alpha}}}

\newcommand{\pos}{\vec{r}}

\newcommand{\Sum}[2]{\sum\limits_{#1}^{#2}}

\let\nablatmp\nabla
\renewcommand{\nabla}{\vec{\nablatmp}}
\DeclarePairedDelimiter\abs{\lvert}{\rvert}
\makeatletter
\let\oldabs\abs
\def\abs{\@ifstar{\oldabs}{\oldabs*}}

\AtBeginDocument{
\heavyrulewidth=.08em
\lightrulewidth=.05em
\cmidrulewidth=.03em
\belowrulesep=.65ex
\belowbottomsep=0pt
\aboverulesep=.4ex
\abovetopsep=0pt
\cmidrulesep=\doublerulesep
\cmidrulekern=.5em
\defaultaddspace=.5em
}
\begin{document}
\title{Constraints on new vector boson mediated electron-nucleus interactions from spectroscopy of polar diatomic molecules}
\date{\today}
\author{Konstantin Gaul}
\email[]{konstantin.gaul@uni-mainz.de}
\author{Lei Cong}
\email[]{conglei1@uni-mainz.de}
\affiliation{Helmholtz Institut Mainz, 55099 Mainz, Germany}
\affiliation{GSI Helmholtzzentrum für Schwerionenforschung GmbH, 64291 Darmstadt, Germany}
\affiliation{Institut f\"ur Physik, Johannes Gutenberg-Universit\"at Mainz, 55099 Mainz, Germany}
\author{Dmitry Budker}
\affiliation{Helmholtz Institut Mainz, 55099 Mainz, Germany}
\affiliation{GSI Helmholtzzentrum für Schwerionenforschung GmbH, 64291 Darmstadt, Germany}
\affiliation{Institut f\"ur Physik, Johannes Gutenberg-Universit\"at Mainz, 55099 Mainz, Germany}
\affiliation{University of California, Berkeley, California 94720, USA}
\begin{abstract}
A measurement of parity violation in the hyperfine structure of
$^{138}$Ba$^{19}$F [E. Altunta\c{s} et al.  Phys. Rev. Lett. 120, 142501
(2018)] is reinterpreted with electronic structure calculations
in terms of beyond Standard Model vector boson mediated electron-nucleus interactions. Our results set constraints on previously unexplored, new boson mediated axial vector-vector nucleus-electron interactions. Similar bounds are obtained by analyzing the atomic parity violation experiment with $^{133}$Cs. Moreover, we show that future
experiments with cold heavy diatomic molecules like $^{225}$RaF can improve the present sensitivity to axial vector-vector nucleus-electron and nucleon-nucleus
interactions by up to five
orders of magnitudes.
\end{abstract}

\maketitle

\emph{Introduction.}---%
The proposal to test for parity ($P$) violation (PV) in weak interactions by Lee and Yang \cite{lee:1956} and its rapid
experimental realization by Wu et al.\ \cite{wu:1957} changed paradigms in the view on
discrete symmetries in fundamental physical processes. In
recent years, in the context of the quest to understand the nature of dark matter, ``new'' bosons mediating interactions that
violate discrete symmetries, in particular parity, have been identified as promising
candidates \cite{bertone:2018,bertone:2005,cong:2024}. Of these, the spin\nobreakdash-1 $Z'$-boson is of particular interest because it is predicted by various extensions of the standard model (SM) of particle physics
\cite{lopez:1997,gomezdumm:1997,rizzo:1998,baek:2006} and its role as a possible link between matter and dark matter \cite{shuve:2014,alves:2014}.

Precision spectroscopy of heavy atoms such as Cs \cite{wood:1997}, Yb
\cite{tsigutkin:2009,antypas:2019}, Tl \cite{vetter:1995} or Dy \cite{nguyen:1997,leefer:2014} serves as a powerful tool to test electroweak PV within the SM as well as PV due to new spin\nobreakdash-1 bosons beyond the SM (BSM)
\cite{dzuba:2017a}. In atoms, nuclear spin-independent effects or more general axial vector (A)-vector (V)
electron (e)-nucleus (N) interactions (AV-eN) dominate. Nuclear spin-dependent PV (NSDPV) effects are usually suppressed. Three interaction types contribute to NSDPV: AV nucleus-electron interactions (AV-Ne), hyperfine interaction induced AV-eN interactions and electromagnetic interactions of electrons with weak nuclear moments like the anapole moment. The latter primarily originate in proton (p) or neutron (n)-nucleus (N) AV interactions (AV-(n,p)N). Dzuba et al. extracted the currently best bounds on new boson mediated AV-eN interactions from the  6s-7s E1 transition in $^{133}$Cs. The corresponding coupling constant was constrained to be $\abs{g^\mathrm{A}_\mathrm{e}g^\mathrm{V}_\mathrm{N}}<3\times10^{-14}/(M^2c^4/\mathrm{eV}^2)$ \cite{dzuba:2017a}. From comparison of the PV effect on hyperfine components of the forbidden atomic transition in $^{133}$Cs they inferred the best constraints on AV-pN interactions to be $\abs{g^\mathrm{A}_\mathrm{p}g^\mathrm{V}_\mathrm{N}}<3\times10^{-7}/(M^2c^4/\mathrm{eV}^2)$. AV-Ne interactions were not discussed in Ref.\,\cite{dzuba:2017a}. In addition, constraints on coupling strength of new vector boson mediated can be obtained from combining bounds on individual coupling strength  $g_\mathrm{e}^\mathrm{V}$ \cite{delaunay:2017}, $g_\mathrm{e}^\mathrm{A}$ \cite{ficek:2017}, $g_\mathrm{p}^\mathrm{A}$ \cite{ramsey:1979}. However, these combined bounds are model dependent and, therefore, PV experiments with atoms and molecules provide complementary information (for a discussion see Ref.\,\cite{cong:2024})

In contrast to atomic PV experiments, in polar diatomic molecules 
AV-eN interactions are usually suppressed, rendering them most
sensitive to NSDPV processes \cite{labzowsky:1978,sushkov:1978,kozlov:1995}. NSDPV manifests in these molecules as a mixing of $\Lambda$- or $\Omega$-doublet states, which are connected by space- and time-parity inversion. In the molecular hyperfine structure, the effective PV spin-rotation (sr) Hamiltonian is \cite{kozlov:1995}:
\begin{equation} 
\hat{H}_\mathrm{PV,sr} =
k_\mathcal{A} W_\mathrm{a}\vec{\lambda}\times\vec{S}'\cdot\vec{I}\,,
\label{eq:Heff_anapole}
\end{equation}
where $\vec{\lambda}$ is the unit vector pointing from the heavy to the light
nucleus, $\vec{S}'$ is the effective electron spin , $\vec{I}$ is
the nuclear spin. In our notations, $\vec{S}'$ and $\vec{I}$ do not contain $\hbar$. Here $W_\mathrm{a}=\Braket{\Omega|\hat{H}_\mathrm{NSDPV} |-\Omega}/\Omega$, where $\hat{H}_\mathrm{NSDPV}$ is the NSDPV-Hamiltonian. $k_\mathcal{A}=k_\mathrm{a}+k_\mathrm{AV} +k_\mathrm{hf}$ is a nuclear structure dependent parameter containing contributions from the nuclear anapole moment (a),
electroweak AV-Ne, 
and hyperfine (hf) induced AV-eN current interactions. 
The effective constant $k_\mathrm{hf}$ depends on the electronic structure. 

Experiments have been proposed for a number of molecular
systems (see reviews \cite{kozlov:1995,khriplovich:2004,titov:2006,roberts:2015,berger:2019} and references therein as well as Refs.\,\cite{demille:2008,isaev:2010,isaev:2012,isaev:2014,borschevsky:2013}). In 2018,
in an experiment with \ce{{}^{138}BaF} ($I_\mathrm{Ba}=0$) a sufficient resolution was 
demonstrated to detect the weak nuclear-spin dependent
$P$-odd contribution from the $^{137}$Ba nucleus if the experiment were
performed with \ce{{}^{137}BaF}\,\cite{altuntas:2018}. Recently new techniques employing Penning trapped molecular ions were proposed
to further improve on this experimental precision \cite{karthein:2024}. 

In this letter we use molecular density functional theory calculations to extract constraints on so far unexplored new boson mediated AV-Ne
interactions from the experiment with $^{138}$Ba$^{19}$F \cite{altuntas:2018}. In addition we extract comparable constraints from comparison of the PV effect on hyperfine components of the forbidden atomic transition in $^{133}$Cs. 
To our best knowledge previously only constraints on AV-ne and AV-pe interaction for new boson masses  smaller than $10^{-4}\,\mathrm{eV}/c^2$ were reported \cite{hunter:2013,wang:2023} (see also Ref.\,\cite{cong:2024} for a review). Our results extend 
to the subatomic scale ($M \sim 10^{5}\,\mathrm{eV}/c^2$).
Finally, we demonstrate that PV experiments with polar diatomic molecules have the potential to improve the present sensitivity to AV-Ne and AV-(n,p)N by several orders of magnitude.

\begin{figure}
\includegraphics[width=.5\textwidth]{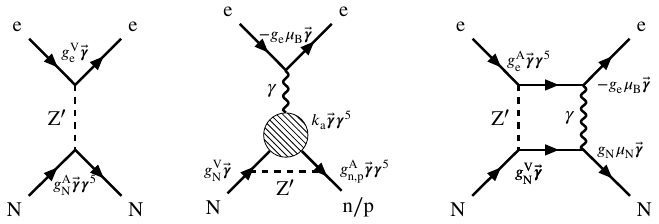}
\caption{Feynman diagrams of contributions to the parity violating mixing in the BaF experiment from new $Z'$-like bosons. The left diagram shows the direct new boson mediated nuclear spin dependent AV-Ne interaction between the nucleus (N) and the electron (e). The middle diagram shows the electromagnetic (mediated by a photon $\gamma$) interaction of the electron with a nuclear anapole moment $k_\mathrm{a}$, which is shown as a big hatched node, which arises due to AV-(n,p)N interactions. The right diagram shows the second order interaction of an electron coupling via a new boson AV-eN interaction and the electromagnetic hyperfine interaction to the nucleus with electron and nuclear $g$-factor being $g_e$,  $g_\mathrm{N}$ respectively and $\mu_\mathrm{B}$, $\mu_\mathrm{N}$ being the Bohr and nuclear magneton.}
\label{fig:FD}
\end{figure}

\emph{Theory.}---%
A new spin\nobreakdash-1 boson would change the PV matrix element $k_\mathcal{A}W_\mathrm{a}$ due to the AV interactions visualized in Fig.\,\ref{fig:FD} by 
\begin{multline}
\Delta k_{\mathcal{A},K} W_{\mathrm{a},K}=g_{\mathrm{N},K}^\mathrm{A}g_\mathrm{e}^\mathrm{V} W_{\mathrm{AV\text{-}Ne},K} +g_{\mathrm{N},K}^\mathrm{V}g_\mathrm{e}^\mathrm{A}W_{\mathrm{AV\text{-}eN},K}\\+ g^\mathrm{A}_\mathrm{n,p}g^\mathrm{V}_\mathrm{N} k_{\mathrm{a,AV\text{-}(n,p)N},K}
\frac{M^2c\,e}{\epsilon_0\hbar^3}\lim\limits_{M\rightarrow\infty}W_{\mathrm{AV\text{-}Ne},K}
\end{multline}
for every spin-carrying nucleus $K$ in the molecule. $g^\mathrm{A}_i$ and $g^\mathrm{V}_i$ are the axial vector
and vector coupling constants of particle $i$, where $i$ is the electron (e), the nucleus (N) or a single nucleon (n,p). For nuclei, the coupling constants are approximately given as effective coupling constants
$g_{\mathrm{N},K}^\mathrm{A}\approx\Braket{\pauli}_{\mathrm{p},K}
g_\mathrm{p}^\mathrm{A} + \Braket{\pauli}_{\mathrm{n},K}
g_\mathrm{n}^\mathrm{A}$ and $g_{\mathrm{N},K}^\mathrm{V}\approx(Z_K
g_\mathrm{p}^\mathrm{V} +N_K g_\mathrm{n}^\mathrm{V})/A_K$, respectively,
where we introduced the average proton and neutron spin,
$\Braket{\pauli}_{\mathrm{p},K}$, $\Braket{\pauli}_{\mathrm{n},K}$, as well as the proton, neutron and mass numbers of nucleus $K$,  denoted as $Z_K$, $N_K$, $A_K$ respectively. 

The three contributions stemming from AV-Ne, AV-(n,p)N and AV-eN interactions arise in analogy to the weak interactions as mixing of $\Omega$ and $-\Omega$ states. The electronic structure coefficients $W$ are computed similarly to SM PV matrix elements as
\begin{align}
W_\text{AV-Ne} &= \Braket{\Omega|\hat{H}_\text{AV-Ne} |-\Omega}/\Omega\,, \\
W_\text{AV-eN} &= \Sum{a\not=\Omega}{}\frac{\Braket{\Omega|\hat{H}_\mathrm{hf}|a}\Braket{a|\hat{H}_\text{AV-eN} |-\Omega}}{\Omega(E_{\Omega}-E_a)} \\\nonumber
&+ \Sum{a\not=\Omega}{}\frac{\Braket{\Omega|\hat{H}_\text{AV-eN}|a}\Braket{a|\hat{H}_\mathrm{hf} |-\Omega}}{\Omega(E_{\Omega}-E_a)}\,,
\end{align}  
where $W_\text{AV-eN}$ is represented here in second order perturbation theory as sum over electronic states $\Ket{a}$ with energy $E_a$ except the $\Ket{\Omega}$ state with energy $E_\Omega$. $\hat{H}_\mathrm{hf}$ is the electromagnetic hyperfine interaction operator. For more details see the Supplemental Material. A new spin\nobreakdash-1 boson could couple a
valence nucleon to the other nucleons and contribute to the nuclear anapole
moment of nucleus $K$ as 
\cite{dzuba:2017a}:
\begin{equation}
k_{\text{a,AV-(n,p)N},K} = \begin{cases}
\frac{18\,e\,R_K}{25\,c}\mu_\mathrm{n,p}\,A_K&M c/\hbar\ll 1/R_K\\
\frac{9\,e\,\hbar^2}{5\,c^3M^2\,R_K}\mu_\mathrm{n,p}\,A_K&M c/\hbar\gg 1/R_K\,,
\end{cases}
\end{equation}
where $c$ is the speed of light in vacuum, $e$ is the elementary electric charge, $\hbar$ is the reduced Planck constant, $R_K\approx1.2\,\mathrm{fm}\,A_K^{1/3}$ is the nuclear charge radius, $\mu_\mathrm{n,p}$ is magnetic dipole moment of the valence neutron or proton ($\mu_\mathrm{p}=2.79\,\mu_\mathrm{N}$ and $\mu_\mathrm{n}=-1.91\,\mu_\mathrm{N}$) in nuclear magnetons $\mu_\mathrm{N}=e\hbar/(2m_\mathrm{p})$ with $m_\mathrm{p}$ being the proton mass.

Assuming the nucleus to be non-relativistic and having a charge density distribution $\rho_K(\vec{r}_i)$, the PV electron-nucleus interaction operator is \cite{dzuba:2017a,fadeev:2019}:
\begin{align}
\hat{H}_\text{AV-eN} &= \frac{\hbar c}{4\pi} \Sum{i}{N_\mathrm{elec}}\Sum{K}{N_\mathrm{nuc}}A_K\diraccontra{5}_i V_{\mathrm{Y},K}(\pos_i)\,,
\label{eq:aven}\\
\hat{H}_\text{AV-Ne} &= \frac{\hbar c}{4\pi} \Sum{i}{N_\mathrm{elec}}\Sum{K}{N_\mathrm{nuc}}\vec{I}_K\cdot\diraca_iV_{\mathrm{Y},K}(\pos_i)\,,
\label{eq:avne}
\end{align}
where the Yukawa-like electron-nucleus potential is 
\begin{equation}
V_{\mathrm{Y},K}(\pos_i)=\int
\mathrm{d}^3r'\,\rho_K(\pos')\frac{\mathrm{exp}(-M c/\hbar
\abs{\pos_i-\pos_K-\pos'})}{\abs{\pos_i-\pos_K-\pos'}}\,,
\end{equation}
the position operator
of the electron $i$ is $\pos_i$, with $r_{iK}=\abs{\pos_i-\pos_K}$, the sums
run over all electrons $i$ and nuclei $K$. $\pos_K$ and $\vec{I}_K$ are the
position and spin of nucleus $K$. Dirac matrices $\diraca,\diraccontra{5}$ are chosen in standard notation as defined in the Supplemental Material.  In the following we assume a
point-like nucleus for which the Yukawa-like potential reduces to
$V_{\mathrm{Y},A}(\pos_i)=\mathrm{exp}(-M c/\hbar r_{iA})/r_{iA}$. For computation of the sensitivity to AV-(n,p)N interactions we assume a finite nucleus that is described by a spherical Gaussian charge density distribution. For this case, the Yukawa-like potential reduces to $\lim\limits_{M\rightarrow\infty}V_{\mathrm{Y},K} = \hbar^2 4\pi \rho_K(r_{iK})/(M\,c)^2$ in the limit of large boson masses.
For heavy nuclei, finite nuclear size effects are expected to be sizable. In the Supplementary material we provide a comparison of the point-like nuclear model with the finite nuclear model for $M\rightarrow\infty$ and $M\rightarrow0$. This suggests that resulting errors are negligible for $M\rightarrow0$, whereas errors are sizeable for $M\rightarrow\infty$, with the largest relative error ($\sim40\,\%$) being found for AV-eN interactions in RaF. This does not significantly influence the discussion and figures shown in the following.

\emph{Molecular density functional theory calculations}---%
In order to obtain constraints on new-boson-mediated interactions from the $^{138}$Ba$^{19}$F experiment and experiments with $^{137}$BaF as well as isotopologues of RaF and SiO$^+$ we compute electronic enhancement factors $W_\mathrm{AV-eN}$ and $W_\mathrm{AV-Ne}$ with molecular density functional theory employing a quasi-relativistic two-component
Hamiltonian within zeroth order regular approximation (ZORA) with the program
developed in Refs.\,\cite{berger:2005,nahrwold:2009,isaev:2010,isaev:2012,gaul:2020,colombojofre:2022,zulch:2022,bruck:2023} as part of the program package \cite{wullen:1998,wullen:2010} which is based on Turbomole \cite{ahlrichs:1989}. We
employed the same methodology as in Ref.\,\cite{gaul:2024a}: a hybrid local density approximation with 50 \% Fock exchange by Becke (BHandH) \cite{dirac:1930,slater:1951,vosko:1980,becke:1993a} and a large atom-centered Gaussian
basis set by Dyall (dyall.cv4z) \cite{dyall:2006} which was augmented with additional tight s- and p-type functions as described in the Supplemental Material. This method showed an excellent agreement with
relativistic coupled cluster calculations of other symmetry violating properties \cite{gaul:2024a}. More computational
details are provided in the Supplementary Material. At this level of theory, we
can reproduce relativistic coupled cluster calculations of the electroweak
PV matrix element due to the anapole moment in $^{137}$Ba $\abs{W_\mathrm{a}}$ ($147.7\,\mathrm{Hz}$)
\cite{hao:2018} very well ($149\,\mathrm{Hz}$). A similarly excellent agreement is found for $\abs{W_\mathrm{a}}$ in $^{225}$RaF ($1731\,\mathrm{Hz}$) with sophisticated coupled
cluster calculations ($1694\,\mathrm{Hz}$) \cite{kudashov:2014,wilkins:2023}. For $^{29}$SiO$^+$ our computed electroweak matrix element
($15\,\mathrm{Hz}$) is in reasonable agreement with four-component coupled
cluster calculations ($13\,\mathrm{Hz}$) \cite{karthein:2024}. We provide a list of matrix elements contributing to $\Delta k_{\mathcal{A},K} W_{\mathrm{a},K}$ for different boson masses for all studied molecules in the Supplementary
Material. 

\begin{figure}[!htbp]
\includegraphics[width=.5\textwidth]{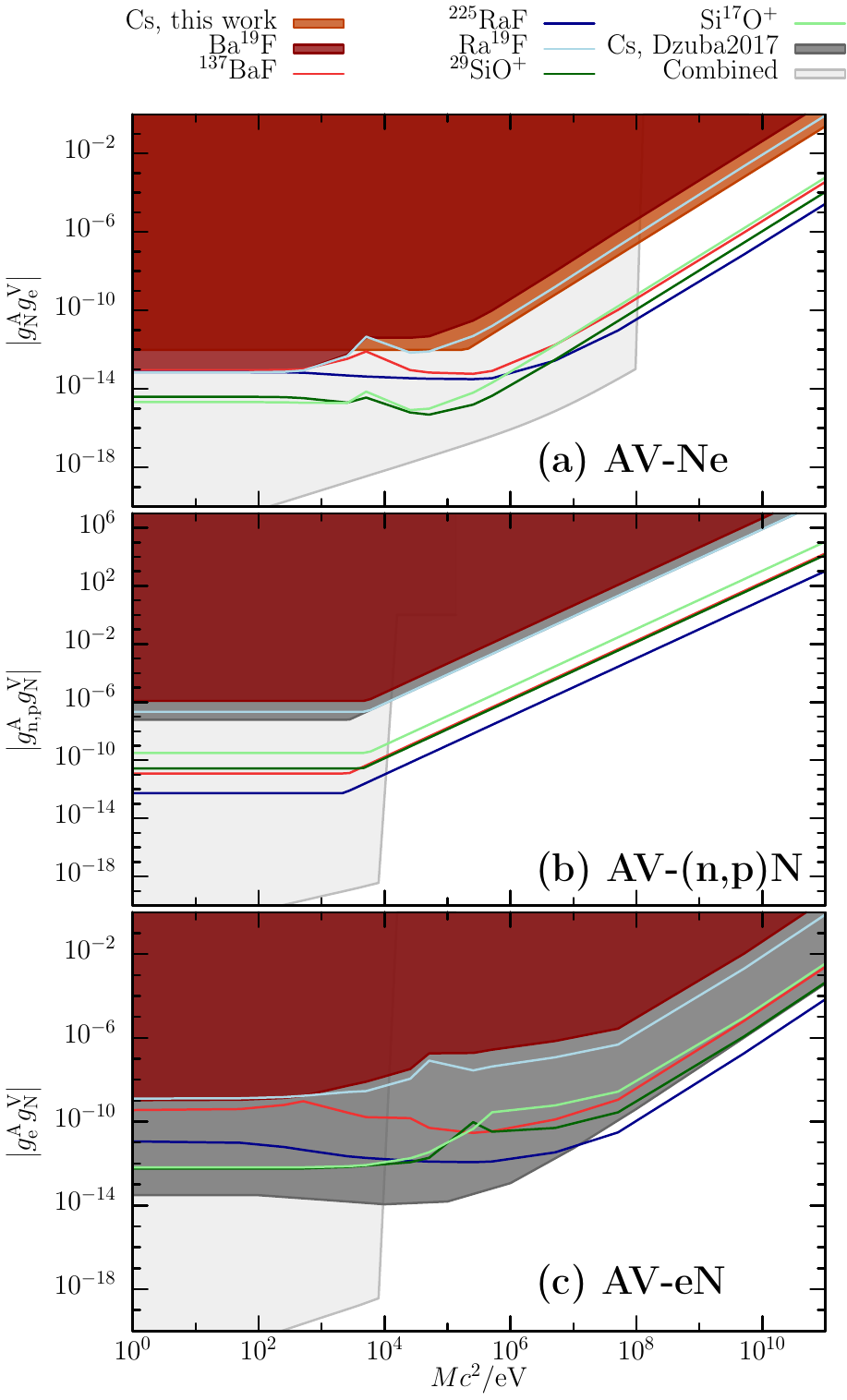}
\caption{Constraints on new boson mediated (a) AV-Ne interactions, (b) AV-(n,p)N interactions and (c) AV-eN interactions from PV experiments with atoms and diatomic molecules. Results of this work from the $^{138}$Ba$^{19}$F experiment and the Cs experiment as well as future experiments with diatomic molecules BaF, RaF and SiO$^+$ are shown. Constraints on AV-(n,p)N and AV-eN interactions are compared to constraints from an atomic PV experiment with Cs obtained in Ref.\,\cite{dzuba:2017a}. For comparison, combined constraints from astrophysical bounds on $g_\mathrm{N}^\mathrm{V}$ \cite{raffelt:1999}, $g_\mathrm{N}^\mathrm{A}$ \cite{dror:2017} and laboratory bounds on $g_\mathrm{e}^\mathrm{V}$ \cite{delaunay:2017}, $g_\mathrm{e}^\mathrm{A}$ \cite{ficek:2017}, $g_\mathrm{p}^\mathrm{A}$ \cite{ramsey:1979} are shown (for discussion see Ref.\,\cite{cong:2024}). The nucleus which interacts with electrons via a new boson is indicated by its explicit mass number. Actual excluded regions are shown as filled areas and projected sensitivities are shown as solid lines. 
}
\label{fig:AV_limits}
\end{figure}

\emph{Constraints from the $^{138}$Ba$^{19}$F experiment}---%
By combining our calculations with the experimental uncertainty of Ref.\,\cite{altuntas:2018}, which was given as $\Delta\nu=0.7\,\mathrm{Hz}$ we extract constraints on new spin\nobreakdash-1 boson AV-Ne, AV-eN and AV-pN
interactions, which are shown in Fig.\,\ref{fig:AV_limits} as red area. We computed the SM-PV matrix element to be $k_{\mathcal{A},\mathrm{{}^{19}F}}\left|W_\mathrm{a,{}^{19}F}\right|=0.014\,h\,\mathrm{Hz}\,k_{\mathcal{A},\mathrm{{}^{19}F}}$. Our prediction of $W_\mathrm{a}$ is on the same order of magnitude as the estimate of Ref.\,\cite{altuntas:2018}. With the estimate $k_{\mathcal{A},\mathrm{{}^{19}F}}\approx-0.08$ \cite{flambaum:1984,demille:2008,altuntas:2018} $k_\mathcal{A}W_\mathrm{a}$ is significantly smaller than the current experimental resolution.

To the best of our knowledge, this is the first direct constraint for AV-Ne interactions [shown in Fig.\,\ref{fig:AV_limits}\,(a)]. For comparison, we extracted constraints on AV-Ne interactions from the atomic Cs PV experiment by relating the uncertainty in
$k_{\mathcal{A},{}^{133}\mathrm{Cs}}$ from Refs.\,\cite{flambaum:1997,dzuba:2017a} $\Delta
k_{\mathcal{A},{}^{133}\mathrm{Cs}}\approx0.236$ to AV-Ne interactions under the assumption that the new boson contribution to the nuclear anapole moment is zero. For $M>\,\mathrm{MeV}/c^2$, the ratio between the electroweak NSDPV matrix element and the new boson NSDPV matrix element is $(\hbar^3\sqrt{2}g_\mathrm{e}^\mathrm{V}g_\mathrm{N}^\mathrm{A})/(4\pi
G_\mathrm{F}cM^2)\approx 9.62
\times 10^{21} g_\mathrm{e}^\mathrm{V}g_\mathrm{N}^\mathrm{A}\,\mathrm{eV}^2/(M^2 c^4)$.  In analogy to new boson-mediated AV-eN
interactions in Cs \cite{dzuba:2017a}, we assume that the limit $M\rightarrow0$ is reached for masses $M\lesssim200\,\mathrm{keV}/c^2$. In this limit, we obtain $\lim\limits_{M\rightarrow0}\Delta k_\mathcal{A}\approx4.81 \times
10^{11} g_\mathrm{e}^\mathrm{V}g_\mathrm{e}^\mathrm{A}$. The obtained constraints
are shown in Fig.\,\ref{fig:AV_limits}\,(a) as the orange area. These Cs constraints are about one order of magnitude weaker than constraints from $^{138}$Ba$^{19}$F for $M<\mathrm{keV}$ and are not more than one order of
magnitude stronger for $M>100\,\mathrm{keV}$, even though the
nuclear charge number of Cs is six times larger than that of F. This highlights the advantage of molecules over atoms in searches
for new boson mediated AV-Ne interactions.

In Fig.\,\ref{fig:AV_limits}\,(b,c), we compare bounds obtained in this work to the existing constraints on AV-(n,p)N and AV-eN interactions, shown as gray area.
There are also constraints on AV-e(n,p) interactions \cite{antypas:2019}, but they are weaker by a few orders of magnitude compared to the constraints on AV-eN \cite{dzuba:2017a}. 
$^{138}$Ba$^{19}$F is only slightly less sensitive to AV-(n,p)N interactions than the Cs experiment \cite{dzuba:2017a}.
As expected from the generally lower sensitivity of $\Ket{\Omega}$ states to nuclear spin-independent PV, discussed in the introduction, the $^{138}$Ba$^{19}$F experiment is less sensitive to AV-eN interaction than atomic PV experiments \cite{dzuba:2017a}.

The sensitivity of the BaF molecular experiments can be improved in a straightforward way \cite{demille:2008,altuntas:2018}.
Assuming that the experiment would have been done with $^{137}$Ba$^{18}$F,
we see [bright red line in Fig.\,\ref{fig:AV_limits} (a),\,(b)] that it would not only improve the sensitivity to AV-eN and AV-Ne interactions by 1–5 orders of magnitude, but also enhance the current sensitivity to AV-(n,p)N interactions by about four orders of magnitude. We note here
that future experiments employing vibrational spectroscopy of chiral molecules
may be more sensitive to AV-(n,p)N interactions for boson masses $M<10^4\,\mathrm{eV}/c^2$
\cite{baruch:2024}.

It
has to be emphasized that the electroweak $P$-odd effect of an odd Ba nucleus
would overshadow the new boson effect and extracting BSM physics will rely on accurate predictions by molecular and nuclear theory. Whereas the accuracy of molecular calculations is currently 1.5\,\% for BaF \cite{hao:2018}, accurate nuclear structure calculations are lacking. Assuming an uncertainty of $\sim30\,\%$ in the nuclear anapole moment calculation of $^{137}$Ba as done for $^{133}$Cs \cite{flambaum:1997}, the uncertainty on $\Delta k_\mathcal{A}W_\mathrm{a}$ may increase roughly to $\sim 3\,\mathrm{Hz}$ for $^{137}$BaF, i.e. by a factor of four. Here we coarsely estimated the size of the nuclear anapole moment as \cite{flambaum:1984,flambaum:1985} $k_\mathrm{a}\sim1.15\times10^{-3}g_\mathrm{n}\mu_\mathrm{n}A^{2/3}(I+1/2)/(I+1)\leq0.07$, where $g_\mathrm{n}=0.9\pm0.6$ \cite{fadeev:2019a}. However, we are confident that a first measurement of PV in BaF will motivate improved calculations of nuclear anapole moments.
The different dependence on $Z$ and $A$ of SM and BSM contributions to $k_\mathcal{A}W_\mathrm{a}$, can facilitate the disentanglement of SM-PV and BSM-PV via
comparison of different isotopologues of BaF. Measurements of stable or long-lived isotopes of ${^{130-141}}$Ba could be compared and would allow to distinguish the
observed parity violating mixing from the SM contribution without
being limited by theory uncertainties. The experimental uncertainty in BaF is expected to be further decreased by direct laser-cooling \cite{kogel:2025}, which was recently proposed for different isotopologues \cite{kogel:2024}.

\emph{Projected sensitivity of PV experiments with other diatomic molecules}---%
The sensitivity of the BaF PV experiment is expected to be improved by future
experiments with RaF by several orders of magnitude
\cite{isaev:2010,Isaev:2013,kudashov:2014} and ongoing experimental efforts
promise a realization of an RaF PV experiment in the near future
\cite{garciaruiz:2020,udrescu:2021,wilkins:2023}. Ra has a long chain of isotopes with a half-life above one second which renders precision spectroscopy of different isotopologues of RaF for separating SM-PV and BSM-PV particularly promising \cite{garciaruiz:2020,udrescu:2021}. For comparison with BaF, we assume the
experimental sensitivity of the emerging RaF experiment to be equivalent to the BaF PV experiment 
\cite{altuntas:2018} ($\Delta\nu=0.7\,\mathrm{Hz}$) and do not include a hypothetical nuclear theory uncertainty. 
Projected sensitivities of RaF are shown as blue lines
in Fig.\,\ref{fig:AV_limits}. The RaF PV experiment is planned to be performed with $^{225}$Ra. We provide for comparison also the projected BSM-PV sensitivity for an experiment that would probe the $^{19}$F nucleus. We find that Ra$^{19}$F is slightly more sensitive to PV than Ba$^{19}$F for masses $M>10^4\,\mathrm{eV}/c^2$. RaF experiments with nonzero-spin Ra nuclei could improve the sensitivity of the BaF experiments to AV-Ne interactions by up to one order of magnitude and to AV-(n,p)N interactions by roughly two orders
of magnitude. Moreover, an equally large improvement of the BaF experiment is expected for AV-eN
interactions. This may allow to improve the sensitivity of PV
experiments to AV-eN interactions for masses $M>10^7\,\mathrm{eV}/c^2$.We note that the experimental
resolution in RaF will likely be better than $0.7\,\mathrm{Hz}$ through direct laser-cooling of RaF \cite{isaev:2010,udrescu:2024,wilkins:2023,gaul:2024b}.

In contrast to experiments that aim for a measurement of electroweak PV, in studies of PV due to new bosons, lighter molecules
have the advantage of small SM-PV effects and therefore may be particularly interesting to search for new bosons with masses $M<10\,\mathrm{keV}$. Recently, systems with light elements
such as MgCN or SiO$^+$ were discussed for electroweak PV experiments in a Penning trap
\cite{hao:2020,karthein:2024}. 
The short-term precision of a PV experiment with SiO$^+$ was
estimated to be limited by roughly $5\,\%$ systematic uncertainty on
the PV matrix-element, which amounts to about $\sim0.02\,\mathrm{Hz}$
\cite{karthein:2024}. Assuming the theoretical prediction of the anapole moment is limited by a smaller uncertainty, constraints shown in
Fig.\,\ref{fig:AV_limits} are extracted with our calculations.
Those systems are competitive to upcoming
experiments with $^{225}$RaF or $^{137}$BaF. Assuming that the uncertainty of calculations of the anapole moment will be considerably smaller than 30\,\%, heavier molecules would, however, show a considerably larger sensitivity to AV-(n,p)N interactions as well as AV-Ne and AV-eN interactions for $M>10^4\,\mathrm{eV}/c^2$. In contrast, for $M<10^4\,\mathrm{eV}/c^2$, the projected sensitivity of SiO$^+$ to AV-Ne and AV-eN is more than one order of magnitude stronger than that obtained for $^{225}$RaF or $^{137}$BaF even if the uncertainty of nuclear structure theory can be ignored. However, for assumed experimental uncertainties atomic experiments are still one to two orders of magnitude more sensitive to AV-eN interactions for $M<10^4\,\mathrm{eV}/c^2$ .

Finally, we note that for extracting rigorous bounds on new bosons all possible sources of PV to an observed PV effect have ultimately to be considered simultaneously, i.e. AV-Ne, AV-eN and AV-(n,p)N interactions within the SM and beyond. This requires to have at least four complementary experiments, which need to be analyzed in a multivariate model similar to beyond Standard model CP-violation \cite{chupp:2015,gaul:2024a,degenkolb:2024}. Therefore, we consider the complementarity of molecular PV experiments to atomic experiments a crucial ingredient to future searches for BSM Z'-like bosons.

\emph{Conclusion}---%
We have analyzed PV experiments with diatomic molecules with respect to their
sensitivity to new spin\nobreakdash-1 bosons. By performing density functional calculations
of new-boson mediated PV electron-nucleus interactions in $^{138}$Ba$^{19}$F we were able to set constraints on previously unexplored, new boson mediated AV-Ne as shown in Fig.\,\ref{fig:AV_limits}\,(a). We could extract similarly tight constraints from the $^{133}$Cs PV experiment.
Moreover, we could demonstrate by calculation of new spin\nobreakdash-1 boson interactions
in the diatomic molecules $^{137}$BaF, RaF and SiO$^+$, that upcoming PV experiments with
diatomic molecules can be used to improve the current sensitivity to AV-Ne and AV-(n,p)N interactions by two to five orders of magnitude when assuming state-of-the-art experimental resolution. Our
calculations suggest that future experiments with heavy diatomic molecules have
the potential to improve the sensitivity over the current atomic PV experiments to AV-eN interactions for boson masses $M>10\,\mathrm{MeV}/c^2$. 

Recent developments in the field include the proposal to measure the nuclear spin-dependent PV
interaction via indirect PV nuclear spin-spin couplings using TlF
\cite{blanchard:2023}. This approach shows promise in enhancing the
sensitivity of molecular experiments. Moreover, recent nuclear magnetic resonance experiments with more complex, chiral molecules \cite{vandyke:2024} have demonstrated an alternative probe of NSDPV. In addition, vibrational spectroscopy of chiral molecules, which is a prospective direct probe of new vector bosons \cite{gaul:2020c,gaul:2020d}, has been shown to provide sensitivity to AV-eN interactions that is competitive to atomic PV experiments \cite{baruch:2024}.

\emph{Acknowledgements}---%
We thank Yotam Soreq for fruitful discussion and David P. DeMille for helpful comments on our manuscript. K.G. is indebted to the Fonds der Chemischen Industrie (FCI) for generous funding through a Liebig fellowship. We gratefully acknowledge computing time at the supercomputer MOGON 2 at Johannes Gutenberg University Mainz (hpc.uni-mainz.de), which is a member of the AHRP (Alliance for High Performance Computing in Rhineland Palatinate,  www.ahrp.info) and the Gauss Alliance e.V. This research was supported in part by the German Research Foundation (DFG), project ID 390831469: EXC 2118 (PRISMA+ Cluster of Excellence) and project ID BU3035/10-1.

\bibliography{bibexport.bib}

\end{document}